\documentclass[%
 reprint, 
 amsmath,amssymb,
 aps, physrev
]{revtex4-2}

\usepackage{graphicx}
\usepackage{dcolumn}
\usepackage{bm}
\usepackage{braket}

\begin{document}


\title{\textbf{
Nearly Transform-Limited Pulse-Like Temporal and Narrowband Spectral Responses in Sum-Frequency Generation from Bright Twin Beams}}%

\author{Yujiro Eto}
\email{eto.yujiro.8r@kyoto-u.ac.jp}
\affiliation{
Frontier Quantum Measurement Unit, Center for Science Adventure and Collaborative Research Advancement (SACRA), Kyoto University, Kyoto 606-8502, Japan
}%

\date{\today}

\begin{abstract}
We simultaneously characterize the temporal and spectral responses of the coherent component in bright twin beams using a spectrally resolved, delay-dependent sum-frequency generation (SFG) measurement.
The measured temporal-correlation width is $0.138~\mathrm{ps}$, close to the transform-limited value of $0.126~\mathrm{ps}$ estimated from the measured twin-beam spectrum. 
The observed spectral-correlation width is $0.176~\mathrm{rad/ps}$ and is limited by the spectrometer resolution. 
The corresponding time–frequency width product is $0.0243$, far below the value of $1$ for SFG between two independent transform-limited Gaussian pulses. 
\end{abstract}

\maketitle

Time–frequency-entangled photon pairs possess strong temporal and spectral correlations and have been widely studied as useful light sources for quantum spectroscopy and metrology~\cite{RevModPhys.88.045008,10.1021/acs.accounts.8b00173}.
For photon pairs generated by parametric down-conversion (PDC), a broad PDC bandwidth can provide a short arrival-time difference, while a narrow pump bandwidth can confine the frequency sum~\cite{PhysRevLett.94.043602,PhysRevLett.94.073601}.
These properties can provide a combination of high temporal and spectral resolution beyond the conventional Fourier constraint for independent transform-limited optical pulses.
Based on these time–frequency correlations, several spectroscopic schemes have been proposed, including time–frequency-resolved Raman and multidimensional fluorescence spectroscopy~\cite{zhang2022entangled,fan2024entangled,doi:10.1126/sciadv.aed7026}.
More recently, time–frequency-resolved fluorescence spectroscopy using correlated photon pairs has been experimentally demonstrated~\cite{alvarez2025correlated}.
Despite this progress, the low photon flux in the photon-pair regime remains a major limitation for nonlinear spectroscopy, where high optical intensities are often required.

To overcome this limitation in optical intensity, the time–frequency correlations of light generated by high-gain PDC, often referred to as bright twin beams or bright squeezed vacuum, can be read out through two-photon processes such as sum-frequency generation (SFG) and two-photon absorption, providing a route to time–frequency spectroscopy at high photon flux. 
SFG is particularly suited to this purpose because advances in nonlinear optical materials and phase-matching techniques enable high conversion efficiency as well as flexible engineering of the operating wavelength and acceptance bandwidth~\cite{SASAKI20001,NIU2022104243}.
In a pioneering study in 2003, Dayan and co-workers reported in a preprint that SFG from broadband squeezed vacuum produces a narrowband coherent component reflecting the pump spectrum, while exhibiting an ultrafast response to variations in the effective relative delay between the signal and idler fields~\cite{dayan2003coherent}.
Dayan later developed a theoretical framework for two-photon interactions with broadband traveling-wave down-converted light, covering the regime from low to high photon flux~\cite{PhysRevA.76.043813}. 
According to this theory, the coherent response arising from correlated signal–idler pairs can simultaneously exhibit an ultrashort temporal response determined by the broad PDC bandwidth and a narrow spectral response determined by the bandwidth of the pump driving the PDC process, while an incoherent background arising from uncorrelated mode combinations is also present.

Recent studies have quantitatively characterized time–frequency correlations in bright squeezed vacuum using a width product related to the Mancini separability criterion~\cite{Cutipa:22}.
In that study, the temporal correlation width was obtained from the second-order correlation function $g^{(2)}(\tau)$, whereas the spectral correlation width was determined from the spectral intensity covariance. 
Their product was shown to lie far below the Fourier-transform constraint.
In our previous work~\cite{Eto:24}, to establish that the measured temporal response directly corresponds to the coherent component arising from correlated signal–idler pairs, we adopted the spectrally resolved SFG approach introduced by Dayan \textit{et al.} in 2003~\cite{dayan2003coherent}, rather than relying on an autocorrelation measurement.
For the same twin-beam source, four-wave-mixing measurements revealed a frequency chirp in the twin-beam correlations, indicating that the temporal response was affected by spectral phase.

In addition to these recent advances, for applications to time-resolved spectroscopy, it is important to determine how closely the temporal correlation of the coherent component approaches the transform limit set by the available PDC bandwidth.
The coherent temporal response is sensitive to refractive-index dispersion and effective dispersion arising from diffraction, either of which can broaden the temporal correlation.
As in conventional ultrafast optics, such temporal broadening directly degrades the achievable temporal resolution and can compromise impulsive excitation.

Here, we use spectrally resolved, delay-dependent SFG measurement to directly evaluate the temporal and spectral responses of the coherent component in bright twin beams.
The spatial modes of the twin beams are predominantly in the fundamental Gaussian mode, which represents a major difference from our previous work~\cite{Eto:24}.
The measured temporal correlation width is \(\sigma^{\mathrm{twin}}_{\tau_-}=0.138~\mathrm{ps}\), close to the transform-limited value of \(0.126~\mathrm{ps}\) estimated from the measured twin-beam spectrum, demonstrating that the coherent temporal response is nearly transform limited. The spectral correlation width obtained from the same SFG measurement is \(\sigma^{\mathrm{twin}}_{\omega_+}=0.176~\mathrm{rad/ps}\), limited by the spectrometer resolution. The resulting width product, \(\sigma^{\mathrm{twin}}_{\tau_-}\sigma^{\mathrm{twin}}_{\omega_+}=0.0243\), is far below the value of 1 for SFG between two independent transform-limited Gaussian pulses under the same width convention. These results show that bright twin beams can simultaneously retain a nearly transform-limited ultrafast temporal response and narrowband spectral selectivity at high photon flux, providing a basis for sub-Fourier time–frequency-resolved spectroscopy.


To quantify the temporal and spectral correlations, we employ the continuous-variable inseparability criterion introduced by Mancini \textit{et al.}~\cite{PhysRevLett.88.120401}.
Denoting the standard deviations of the signal--idler arrival-time difference $t_s-t_i$ and sum angular frequency $\omega_s+\omega_i$ as $\sigma_{\tau_-}$ and $\sigma_{\omega_+}$, respectively, the condition for time--frequency inseparability is given by~\cite{PhysRevLett.120.053601}
\begin{equation}
\sigma_{\tau_-}\sigma_{\omega_+} < 1.
\label{eq:inseparability}
\end{equation}

We consider the classical physical interpretation of this boundary.
The boundary $\sigma_{\tau_-}\sigma_{\omega_+} = 1$ coincides with that obtained for SFG between two independent, identical, transform-limited Gaussian pulses.
Let $\sigma$ denote the spectral standard deviation of each pulse, $T$ their relative delay, and $\omega_{\mathrm{SFG}}$ the SFG angular frequency.
For such pulses, the SFG spectral intensity profile is
\begin{equation} 
S_{\mathrm{SFG}}(\omega_{\mathrm{SFG}}, T) \propto \exp\left( -\frac{\omega_{\mathrm{SFG}}^2}{4\sigma^2} \right) \exp\left( -\sigma^2 T^2 \right).
\end{equation}

From this intensity distribution, the standard deviation of the SFG spectrum and that of the delay-dependent SFG intensity are evaluated as
$\sigma^{\mathrm{gau}}_{\omega_+} = \sqrt{2}\sigma$,
$\sigma^{\mathrm{gau}}_{\tau_-} = \frac{1}{\sqrt{2}\sigma},$
respectively.
Their product consequently satisfies
$\sigma^{\mathrm{gau}}_{\tau_-} \sigma^{\mathrm{gau}}_{\omega_+} = 1.$

Thus, the boundary of the Mancini inseparability criterion ($\sigma_{\tau_-}\sigma_{\omega_+} = 1$) corresponds directly to the product of the temporal and spectral correlation widths at the classical Fourier limit for SFG between independent transform-limited Gaussian pulses.
Fulfilling $\sigma_{\tau_-}\sigma_{\omega_+} < 1$ therefore demonstrates a sub-Fourier time--frequency response relative to this classical SFG benchmark.

\begin{figure}
\centering
\includegraphics[width=0.4\textwidth]{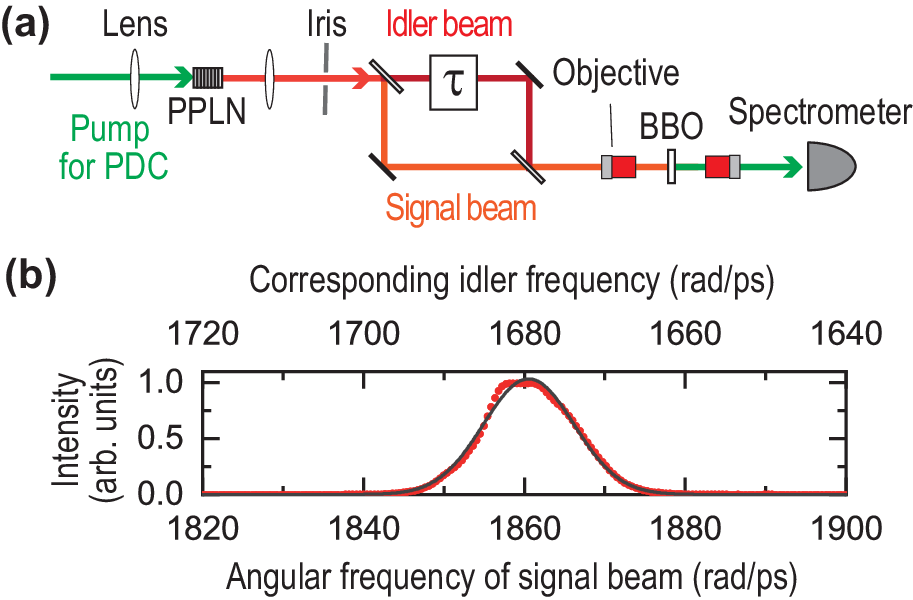}
\caption{
Experimental setup and twin-beam signal spectrum.
(a) Schematic of the experimental setup for twin-beam generation and SFG measurement.
(b) Measured spectrum of the signal beam after spatial filtering by the iris.
The solid curve shows a Gaussian fit to the measured spectrum.
}
\label{fig1}
\end{figure}

The experimental configuration used for twin-beam generation and SFG measurement was similar to that employed in our previous work~\cite{Eto:24}. 
The fundamental source was a $Q$-switched diode-pumped laser (Cobolt, Tor) operating at $1064~\mathrm{nm}$ with a pulse duration of $3.9~\mathrm{ns}$ and a repetition rate of $7~\mathrm{kHz}$. 
Its second harmonic was used as the pump for PDC. 

The experimental arrangement following the PDC stage is shown schematically in Fig.~\ref{fig1}(a).
The pump beam was focused into a $2.5$-cm-long periodically poled LiNbO$_3$ (PPLN) crystal (HC Photonics, SHVIS-SB-25). 
The generated signal and idler beams were collimated with a lens of focal length $f=50~\mathrm{mm}$. 
To obtain a predominantly fundamental Gaussian mode, the transverse wave-vector bandwidth of the twin beams was restricted using an iris with a $1.5$-mm aperture placed at a distance of $2f$ from the exit surface of the PPLN crystal \cite{Eto2026}. 
Figure~\ref{fig1}(b) shows the measured spectrum of the signal beam passed through the iris. 
The solid curve represents a Gaussian fit to the spectrum. 
From the fit, the central angular frequency and spectral standard deviation were estimated to be $1861~\mathrm{rad/ps}$ (corresponding to a wavelength of $1012~\mathrm{nm}$) and $5.6~\mathrm{rad/ps}$, respectively.
The total average power of the twin beams after the iris was 0.67 mW.

The signal and idler beams were then spatially separated by a dichroic mirror, and a variable delay line in the idler path was used to control
their relative delay $\tau$. The two beams were subsequently recombined and focused with an objective lens (M Plan APO NIR $50\times$) into a
$100~\mu\mathrm{m}$-long Type-I BBO crystal (Castech Inc.).
The BBO crystal was mounted on a translation stage for $z$-axis positioning. 
To mitigate Gouy-phase-induced effects and increase the relative contribution of the coherent SFG component, the crystal was displaced by $\sim 200~\mu\mathrm{m}$ from the focus toward the output objective \cite{Eto2026}.
The generated SFG light was collimated with an objective lens (M Plan APO NIR B $20\times$) and analyzed using a high-resolution spectrometer (Andor Technology, Shamrock 500i,
DU970P-BVF).

\begin{figure}
\centering
\includegraphics[width=0.4\textwidth]{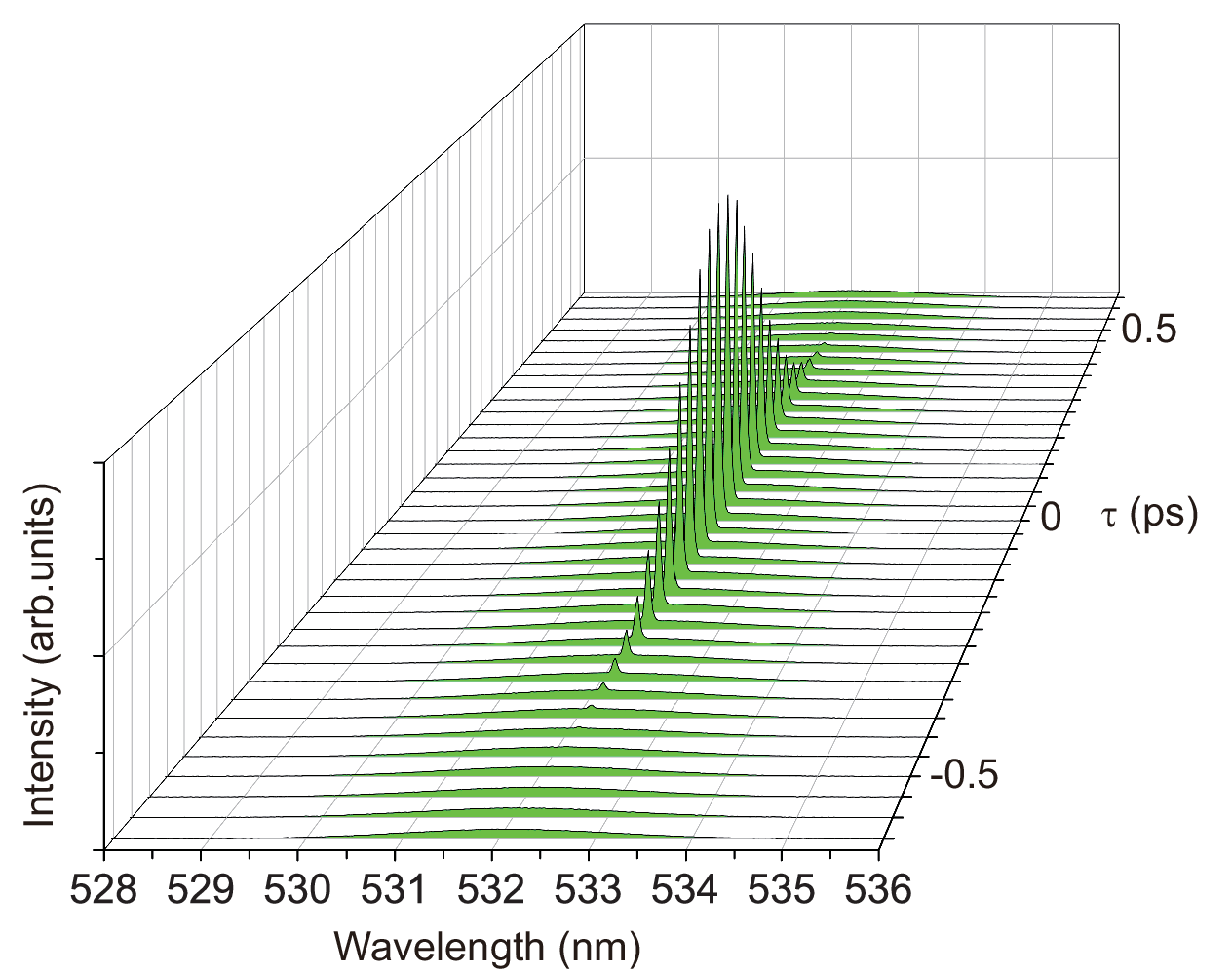}
\caption{
Delay-dependent SFG spectra of the twin beams.
The SFG spectra were measured while scanning the relative time delay $\tau$ between the signal and idler beams.
A sharp spectral peak near 532.17 nm corresponds to the coherent SFG component arising from frequency-correlated signal--idler modes, whereas the broad spectral contribution observed over the full delay range corresponds to the incoherent SFG background generated by spectrally uncorrelated mode combinations.
The coherent component appears only within a finite range of $\tau$, reflecting the temporal correlation of the twin beams.
}
\label{fig2}
\end{figure}

Figure~\ref{fig2} presents the main experimental data of this study, showing the SFG spectra measured while scanning the relative time delay $\tau$. 
A sharp central peak is observed near $532.17~\mathrm{nm}$. 
This peak corresponds to the coherent SFG contribution, in which the spectrum of the PDC pump is reconstructed through the frequency correlation between the signal and idler modes. 
The coherent contribution appears only within a finite range of $\tau$, and its delay dependence therefore reflects the temporal correlation of the twin beams. 
In addition to the sharp central peak, a broad spectral component is faintly visible at each delay.
This broadband contribution corresponds to the incoherent SFG background generated by interactions between spectrally uncorrelated modes.

\begin{figure}[h]
\centering
\includegraphics[width=0.4\textwidth]{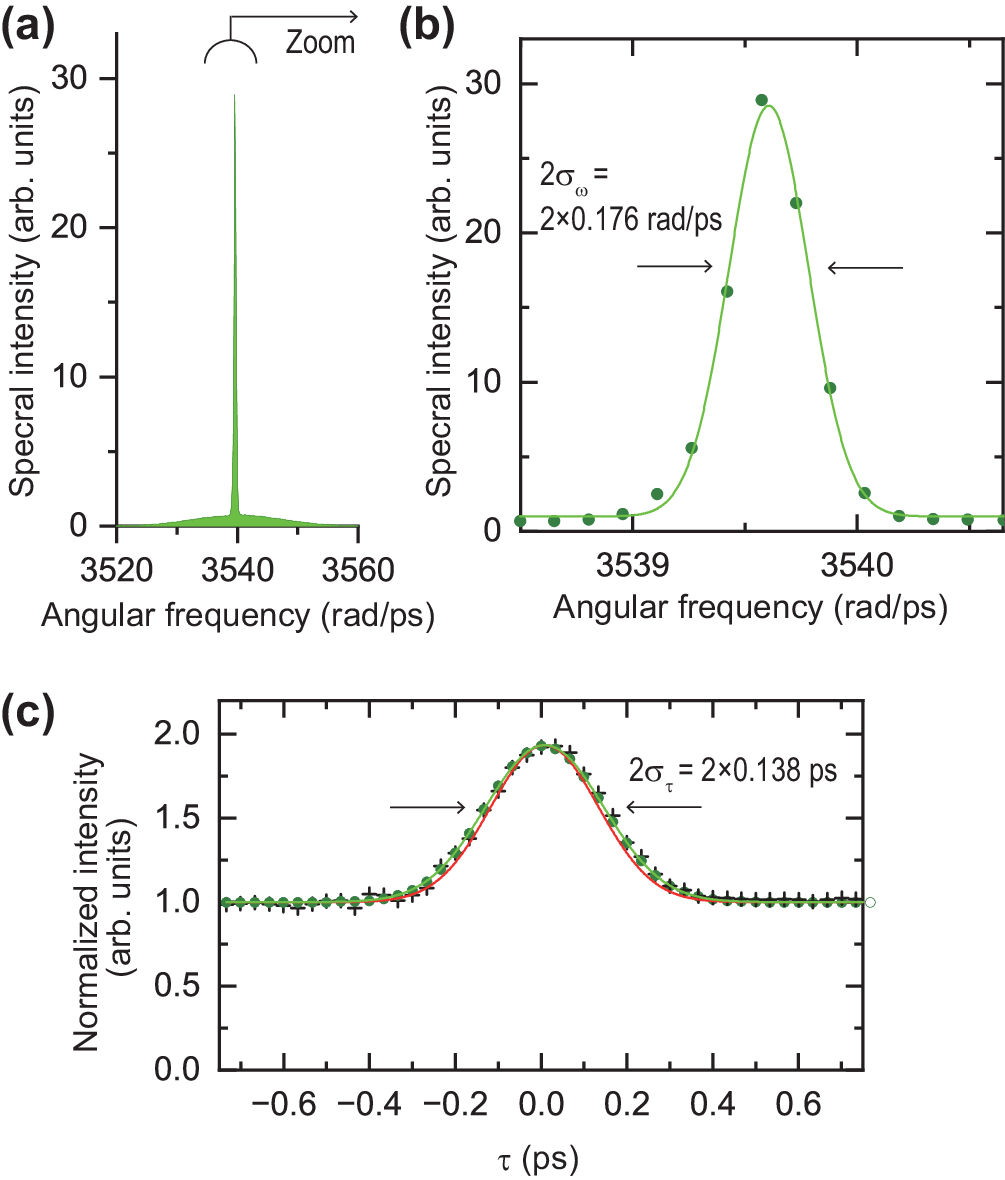}
\caption{
Evaluation of the frequency- and temporal-correlation widths.
(a)  SFG spectrum measured at $\tau=0$, plotted as a function of angular frequency.
(b)  Enlarged view of the SFG spectral profile. 
The solid curve represents a Gaussian fit, yielding a frequency-correlation width of $\sigma^{\mathrm{twin}}_{\omega_+}=0.176~\mathrm{rad/ps}$.
(c)  Integrated SFG intensity as a function of the relative time delay $\tau$. 
The circles show the SFG intensity normalized by the incoherent background measured individually at each delay, 
whereas the crosses represent the intensity normalized by the average value of the incoherent SFG background measured at $\tau = -0.73~\mathrm{ps}$ and $0.9~\mathrm{ps}$.
The solid curve represents a Gaussian fit to the temporal correlation profile, yielding a temporal-correlation width of $\sigma^{\mathrm{twin}}_{\tau_-}=0.138~\mathrm{ps}$.
}
\label{fig3}
\end{figure}

Figure~\ref{fig3}(a) shows the SFG spectrum at $\tau=0$ as a function of angular frequency, which was used to evaluate the frequency-correlation width. 
Figure~\ref{fig3}(b) shows an enlarged view of the sharp coherent spectral profile. 
By fitting this profile with a Gaussian function, the frequency-correlation width was evaluated as $\sigma^{\mathrm{twin}}_{\omega_+}=0.176~\mathrm{rad/ps}$.
This quantity represents the twin-beam counterpart of $\sigma^{\mathrm{gau}}_{\omega_+}$.
This width is comparable to the PDC pump bandwidth measured with the same spectrometer, both being close to the spectral resolution limit of the system.

Figure~\ref{fig3}(c) shows the SFG intensity as a function of the relative time delay $\tau$, obtained by integrating the spectral intensity at each delay in Fig.~\ref{fig2}, which was used to evaluate the temporal-correlation width.
The circles represent the SFG intensity normalized by the incoherent SFG background measured individually at each value of $\tau$, whereas the crosses represent the intensity normalized by the average value of the incoherent SFG background measured at $\tau = -0.73~\mathrm{ps}$ and $0.9~\mathrm{ps}$.
The two profiles agree well, indicating that the incoherent SFG background is nearly independent of $\tau$.
By fitting the temporal-correlation profile with a Gaussian function, the temporal-correlation width was evaluated as $\sigma^{\mathrm{twin}}_{\tau_-}=0.138~\mathrm{ps}$ [green curve in Fig.~\ref{fig3}(c)].
This quantity represents the twin-beam counterpart of $\sigma^{\mathrm{gau}}_{\tau_-}$.

The transform-limited temporal-correlation profile was calculated from the measured twin-beam signal spectrum following the model used in our previous work~\cite{Eto:24}.
The coherent contribution is expressed as
\begin{equation}
G_{\rm si}^{(2)}(\tau) \propto \left| \int H(\omega)\mu^2(\omega) e^{-i\omega\tau} \, d\omega \right|^2,
\label{eq:temporal_correlation}
\end{equation}
where $\mu^2(\omega)$ represents the spectral distribution of the twin-beam signal component and $H(\omega)$ is the phase factor accumulated by the frequency-correlated signal--idler modes~\cite{Eto:24}.
To obtain the transform-limited temporal response, spectral phase dispersion was neglected by setting $H(\omega) = 1$.
The experimentally measured twin-beam signal spectrum shown in Fig.~\ref{fig1}(b) was used for $\mu^2(\omega)$.
The resulting temporal-correlation profile is represented by the red curve in Fig.~\ref{fig3}(c), from which the transform-limited temporal-correlation width was evaluated to be $0.126~\mathrm{ps}$.
The close agreement between the measured ($0.138~\mathrm{ps}$) and transform-limited ($0.126~\mathrm{ps}$) values indicates that the temporal correlation is nearly transform limited, with only minor residual temporal broadening.

Using the frequency- and temporal-correlation widths obtained above, their product is 
\begin{equation}
\sigma^{\mathrm{twin}}_{\tau_-}\sigma^{\mathrm{twin}}_{\omega_+} = 0.138 \times 0.176 = 0.0243.
\end{equation}
This value is well below the classical SFG benchmark of unity obtained for independent transform-limited Gaussian pulses.
The correlation widths used in this evaluation are those of the coherent SFG component.
Therefore, the resulting value represents the idealized time--frequency resolution obtained when the influence of the incoherent SFG background is excluded.

We demonstrated the simultaneous evaluation of nearly transform-limited ultrashort temporal and narrowband spectral responses in bright twin beams generated by high-gain parametric down-conversion. 
Using spectrally resolved, delay-dependent SFG measurements, we evaluated both correlation widths from the same data set, directly establishing their coexistence within the same coherent signal--idler contribution.
The evaluated correlation widths, $\sigma^{\mathrm{twin}}_{\tau_-} = 0.138~\mathrm{ps}$ and $\sigma^{\mathrm{twin}}_{\omega_+} = 0.176~\mathrm{rad/ps}$, yield a time--frequency product of $\sigma^{\mathrm{twin}}_{\tau_-}\sigma^{\mathrm{twin}}_{\omega_+} = 0.0243$, well below the value of 1 expected for SFG between two independent transform-limited Gaussian pulses.
These findings indicate that, if the incoherent component can be effectively suppressed or removed through post-processing, bright twin beams could provide sub-Fourier time--frequency functionality comparable to that of low-gain entangled photon pairs, but at high photon flux.
This work opens up promising avenues for practical applications in high-signal-level quantum spectroscopy and nonlinear optical sensing using bright twin beams with SFG readout.

\paragraph*{Aknowledgments}
We would like to express our gratitude to the HONMAMON co-creative research program between HORIBA and Kyoto University.
This work was supported by the Japan Science and Technology Agency (JST) PRESTO (Grant No. JPMJPR2352), the Japan Society for the Promotion of Science (JSPS) KAKENHI (Grant Nos. 26K21749, 26K22832 and 24K06919), and the RIKEN Transformative Research Innovation Platform (TRIP) initiative. 

\paragraph*{Data availability}
The data are available from the authors upon reasonable request.

\bibliography{apssamp}

\end{document}